\documentclass[twocolumn,epsf,epsfig,showpacs,prl]{revtex4}
\usepackage{graphicx}
\usepackage{dcolumn}
\usepackage{bm}

\begin{document}

\title{Ultrahigh energy cosmic rays as heavy nuclei from cluster accretion shocks}
\author{Susumu Inoue$^{a}$, G{\"u}nter Sigl$^{b,c}$, Francesco Miniati$^d$, and Eric Armengaud$^{e}$}

\affiliation{$^a$ National Astronomical Observatory of Japan, Mitaka, Tokyo 181-8588, Japan}
\affiliation{$^b$APC (AstroParticules et Cosmologie),
11, place Marcelin Berthelot, F-75005 Paris, France}
\affiliation{$^c$ GReCO, IAP, C.N.R.S., 98 bis boulevard Arago, F-75014 Paris, France}
\affiliation{$^d$ Physics Department, Wolfgang-Strasse 16, ETH Z\"urich, 8093 Z\"urich, Switzerland}
\affiliation{$^e$ DSM/DAPNIA, CEA/Saclay, F-91191, Gif-sur-Yvette, France}

\begin{abstract}
Large-scale accretion shocks around massive clusters of galaxies,
generically expected in the cold dark matter scenario of cosmological structure formation,
are shown to be plausible sources of the observed ultrahigh energy cosmic rays (UHECRs)
by accelerating a mixture of heavy nuclei including the iron group elements.
Current observations can be explained if the source composition at injection for the heavier nuclei
is somewhat enhanced from simple expectations for the accreting gas.
The proposed picture should be clearly testable by current and upcoming facilities in the near future
through characteristic features in the UHECR spectrum, composition and anisotropy,
in particular the rapid increase of the average mass composition with energy
from $10^{19}$ to $10^{20}$ eV.
\end{abstract}


\maketitle

{\it Introduction.}
Several decades after their discovery,
the origin of UHECRs with energies $10^{18}$-$10^{20}$ eV and above
remains one of the biggest mysteries in physics and astrophysics \cite{NW00,Hil84}.
The observed global isotropy in the arrival directions strongly suggests
that they are of extragalactic origin.
However, no unambiguous identification with any type of astrophysical source
has been achieved so far.

If UHECRs are protons, photopion interactions with the cosmic microwave background (CMB)
should induce severe energy losses at $\gtrsim 7 \times 10^{19}$ eV
for propagation lengths $\gtrsim 100$ Mpc \cite{GZK66}.
It is also possible that UHECRs are dominated by heavy nuclei, particularly at the highest energies
where the mass composition is observationally quite uncertain \cite{Wat04}.
In this case, photodisintegration and photopair interactions with
the far-infrared background (FIRB) and the CMB govern the energy loss length \cite{PSB76},
which is $\sim 300$ Mpc at $10^{20}$ eV for iron nuclei,
somewhat larger than that for protons \cite{SS99}.

Only a few types of astrophysical objects appear capable of accelerating UHECRs
up to the highest observed energies,
such as the jets of radio-loud active galactic nuclei (AGNs)
and gamma-ray bursts (GRBs) \cite{Hil84}.
Cluster accretion shocks have also been proposed as UHECR sources \cite{NMA95}.
In the currently favored hierarchical scenarios of cosmological structure formation,
massive clusters of galaxies
should be surrounded by highly supersonic accretion flows
that give rise to powerful and long-lived shock waves extending over Mpc scales.
Numerical simulations show that
such shocks should possess high Mach numbers \cite{Min00},
implying that they can accelerate particles with hard spectra at high efficiency
through the first order Fermi mechanism \cite{BE87}.
However, the maximum energy attainable for protons
seem to fall short of $10^{20}$ eV by 1-2 orders of magnitude \cite{KRJ96,KRB97,IAS05}.

In this {\it letter} we show that cluster accretion shocks
may provide a viable explanation of the observed properties of UHECRs
if they accelerate a suitably mixed composition of heavy nuclei,
in particular Fe nuclei up to energies $\sim 10^{20}$ eV. 
Note that ``accretion shocks'' here signify
not only shocks due to infall of cold gas from the diffuse intergalactic medium,
but also those associated with gas inflow along large-scale filaments
that are important in terms of energy dissipation \cite{Min00}.
Our findings hold for plausible assumptions on the source properties 
and a wide range of extragalactic magnetic field strengths,
and should be clearly testable in the near future.

\begin{figure}[ht]
\includegraphics[width=0.41\textwidth]{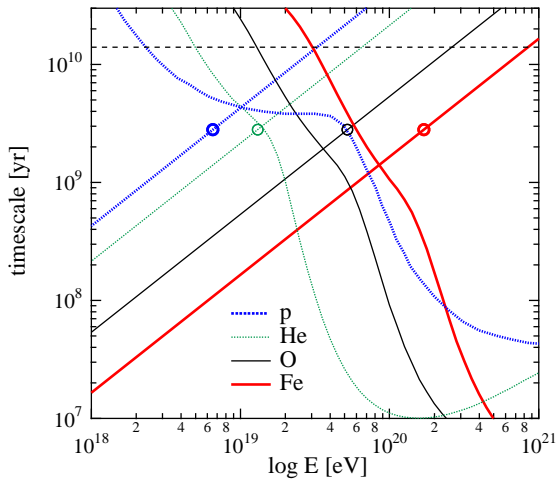}
\caption{Comparison of timescales versus particle energy $E$ at cluster accretion 
shocks for shock acceleration $t_{\rm acc}$ (diagnonal lines),
and for energy losses from interactions with background radiation fields (curves),
for protons (thick dotted), He (thin dotted), O (thin solid) and Fe nuclei (thick solid).
Also indicated are the Hubble time $t_H$ (dashed line)
and the escape-limited maximum energies (circles).}
\label{fig:tacc}
\end{figure}

{\it Model.}
We adopt the cosmological parameters
$h=0.7$, $\Omega_m=0.3$, $\Omega_\Lambda=0.7$ and $\sigma_8=0.9$.
First we estimate the maximum energy 
by considering a fiducial, Coma-like cluster of total mass $M=2 \times 10^{15} M_\odot$,
and extending the discussion of Ref. \cite{KRB97} to include heavy nuclei.
The accretion shock radius is taken to be the virial radius,  $R_s \simeq 3.2$ Mpc,
so that the shock velocity relative to the upstream gas is
$V_s=(4/3)(GM/R_s)^{1/2} \simeq 2200$ km/s \cite{IAS05}.
We choose $B_s=1 \mu$G for the magnetic field at the shock,
as suggested by some recent observations around the virial radius \cite{JH04}. 
The timescale for shock acceleration of particles with energy $E$ and charge $Z$
is $t_{\rm acc}=20 \kappa(E)/V_s^2 = (20/3) (E c/Z e B_s V_s^2)$,
assuming a parallel shock
\footnote{The background magnetic field is parallel to the shock normal.
For oblique shocks with more general magnetic field angles,
$t_{\rm acc}$ can be shorter by up to a factor of 5 or more \cite{Jok87,KRB97}.},
together with the Bohm limit for the diffusion coefficient $\kappa(E)$
which can be induced by CR wave-excitation \cite{Bel04}
and is compatible with supernova remnant (SNR) observations \cite{Par06}. 
The maximum energy $E_{\max}$ can be estimated by comparing $t_{\rm acc}$
with timescales for limiting processes, such as energy loss by
photopair and photopion interactions with the CMB for protons,
and additionally photodisintegration interactions with the FIRB and CMB for nuclei, as defined in Ref.\ \cite{SS99}.
Another limit may be the diffusive escape time from the acceleration region,
$t_{\rm esc} \sim R_s^2/5 \kappa(E)$ \cite{RB93},
which we can equate
with $t_{\rm acc}$ to obtain $E_{\max}/Z \sim (3e B_s R_s V_s/10 c) \simeq 6.5 \times 10^{18}$ eV.
Finally, the Hubble time $t_H = 1.4 \times 10^{10}$ yr represents an absolute upper limit.
These comparisons are illustrated for selected species in Fig.\ref{fig:tacc}.
For protons, $E_{\max} \lesssim 10^{19}$ eV,
limited either by photopair losses or escape, confirming previous findings.
However, heavy nuclei with higher $Z$ and correspondingly shorter $t_{\rm acc}$
can be accelerated to higher energies,
with Fe reaching $\sim 10^{20}$ eV in the same conditions,
notwithstanding photonuclear interaction losses.

For clusters of mass $M$, the rate of gas kinetic energy flow through accretion shocks can be estimated as
$L_{\rm acc} \simeq 9 \times 10^{45} (M/{10^{15} M_\odot})^{5/3} {\rm erg\ s^{-1}}$ \cite{IAS05,Min00}.
We take the number density of clusters with $M \gtrsim 10^{15} M_\odot$ to be
$n_s=2 \times 10^{-6} {\rm Mpc^{-3}}$, in accord with that  
observed locally within 120 Mpc \cite{RB02}, 
and consistent with theoretical mass functions to within a factor of 2 \cite{ST99}.
The power density is $P_{\rm acc} \sim L_{\rm acc} n_s = 2 \times 10^{40} {\rm erg\ s^{-1} Mpc^{-3}}$.
Around $10^{19}$ eV, protons should dominate with propagation lifetime
$\simeq 4 \times 10^9$ yr.
To account for the observed energy density of UHECRs $\gtrsim10^{19}$ eV
of $\sim 10^{-19} {\rm erg \ cm^{-3}}$,
the required power density is then $P_{>19} \sim 3 \times 10^{37} {\rm erg\ s^{-1} Mpc^{-3}}$,
neglecting the effect of magnetic fields.
Thus cluster accretion shocks should reasonably accommodate the energetics of UHECRs.

For a quantitative study of this scenario, we simulate trajectories of nuclei above $10^{19}\,$eV,
accounting for all relevant energy losses and deflections by extragalactic magnetic fields (EGMF),
including secondary nuclei arising from photodisintegration \cite{ASM05,BILS02}.
Since the true nature of EGMF is currently very uncertain \cite{Kro94},
we consider different cases that may bracket the range of possibilities, specifically,
models of relatively strong EGMF that trace large-scale structure as in Refs. \cite{SME04,ASM05},
as well as the case of no EGMF, more in line with some other models \cite{Dol05}.
The effect of Galactic magnetic fields \cite{TYS06} are not included, however.
The observer is chosen in a low EGMF region with some resemblance to the Local Group's actual environment.
Assuming $n_s$ and $P_{\rm acc}$ as above,
we simulate a large number of realizations
where the locations of discrete sources follow the baryon density,
and obtain the average and cosmic variance of the spectra and composition \cite{SME04,ASM05}.

A fraction $f_{\rm CR}$ of the accretion luminosity $L_{\rm acc}$ is converted to UHECRs
with energy distributions $\propto E^{-\alpha} \exp (-E/E_{\max})$,
with $E_{\max}$ and normalization being different for each species.
We set $E_{\max}/Z=5 \times 10^{18}$ eV, which is a fair approximation
to estimates obtained as in Fig.\ref{fig:tacc}.
The escape of UHECRs into intergalactic space is assumed to be energy-independent,
at least in the limited energy range $10^{19}$-$3 \times 10^{20}$ eV.
This could be the case for diffusive escape in directions away from the filaments,
or possibly advective escape during merging events.

For the elemental composition at injection,
we follow Ref.\ \cite{SA05} in taking the abundance ratio by number of He to protons to be 0.042.
All heavier elements are assumed to have the same relative abundances at fixed energy per nucleon
as that inferred for the sources of Galactic cosmic rays at GeV energies \cite{DT96,APO05},
and scaled with respect to protons
by the metallicity of the accreting gas $\zeta$ relative to the solar abundance.
We take $\zeta$=0.2 as suggested by both observations \cite{DM01} and theoretical simulations \cite{CO06} for the warm gas falling in from filaments.
This assumes that particle injection from the thermal gas at cluster accretion shocks 
works in a way similar to Galactic cosmic ray sources, presumably SNRs,
which is plausible considering that the shock velocities are of the same order,
and the temperature of the accreting preshock gas at $10^{5}$-$10^{6}$ K
is that of the hot phase of the interstellar medium.
On the other hand, the nonlinear modification of shock structure by CR pressure
in the upstream region \cite{MD01}
can be stronger in cluster shocks than in SNRs
owing to the acceleration to much higher energies,
and this may possibly lead to a further enhancement of heavier nuclei at injection \cite{BK99}.
We account for this by an additional factor $A^{\beta}$ in the injected abundance of nuclei with atomic number $A$.
In such conditions, progressively heavier nuclei
reach higher $E_{\max}$ and become the dominant species in the source spectrum,
analogous to that inferred for Galactic cosmic ray sources in the knee region \cite{Hil84}.

\begin{figure}[ht]
\includegraphics[width=0.45\textwidth]{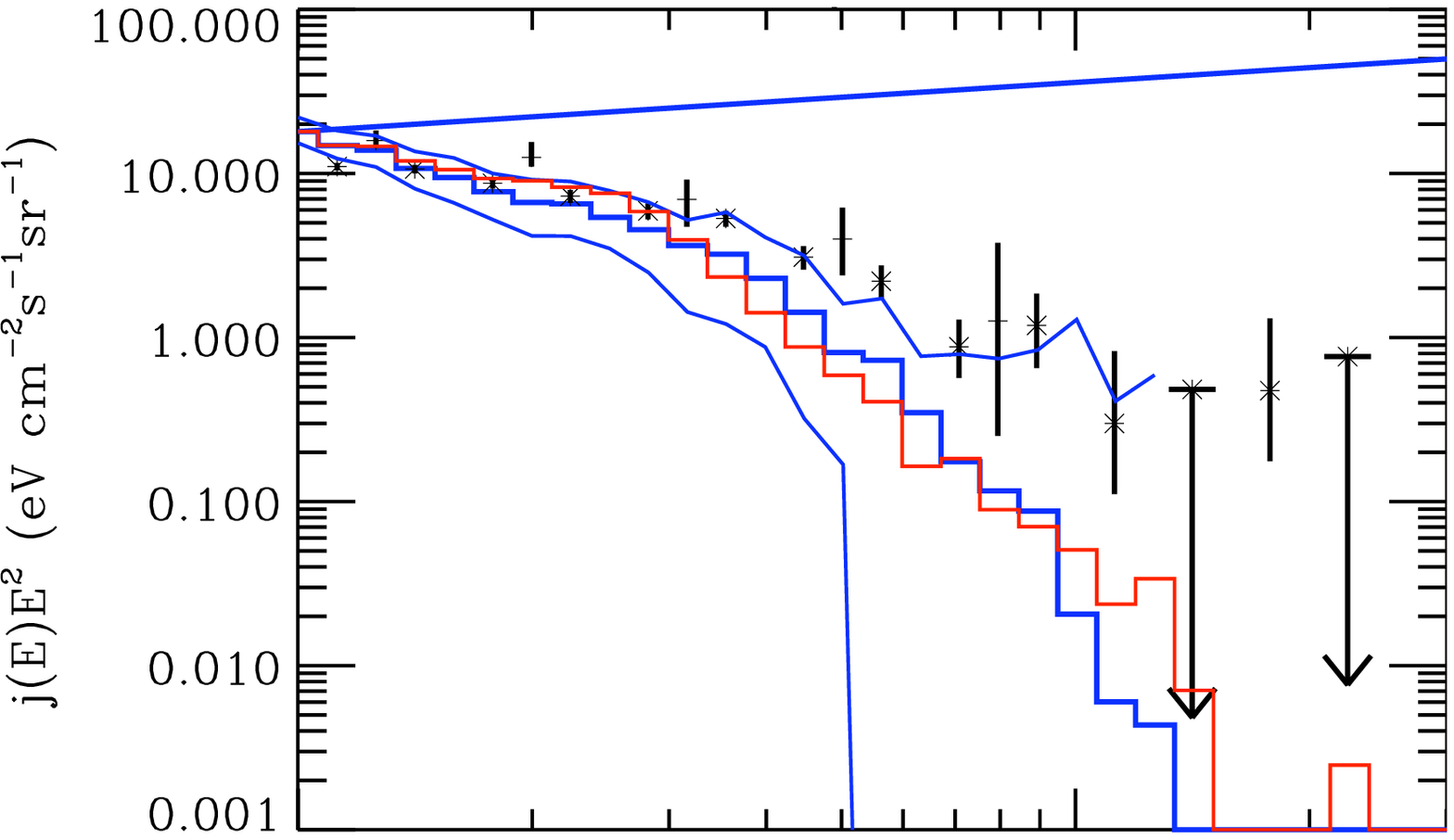}
\includegraphics[width=0.45\textwidth]{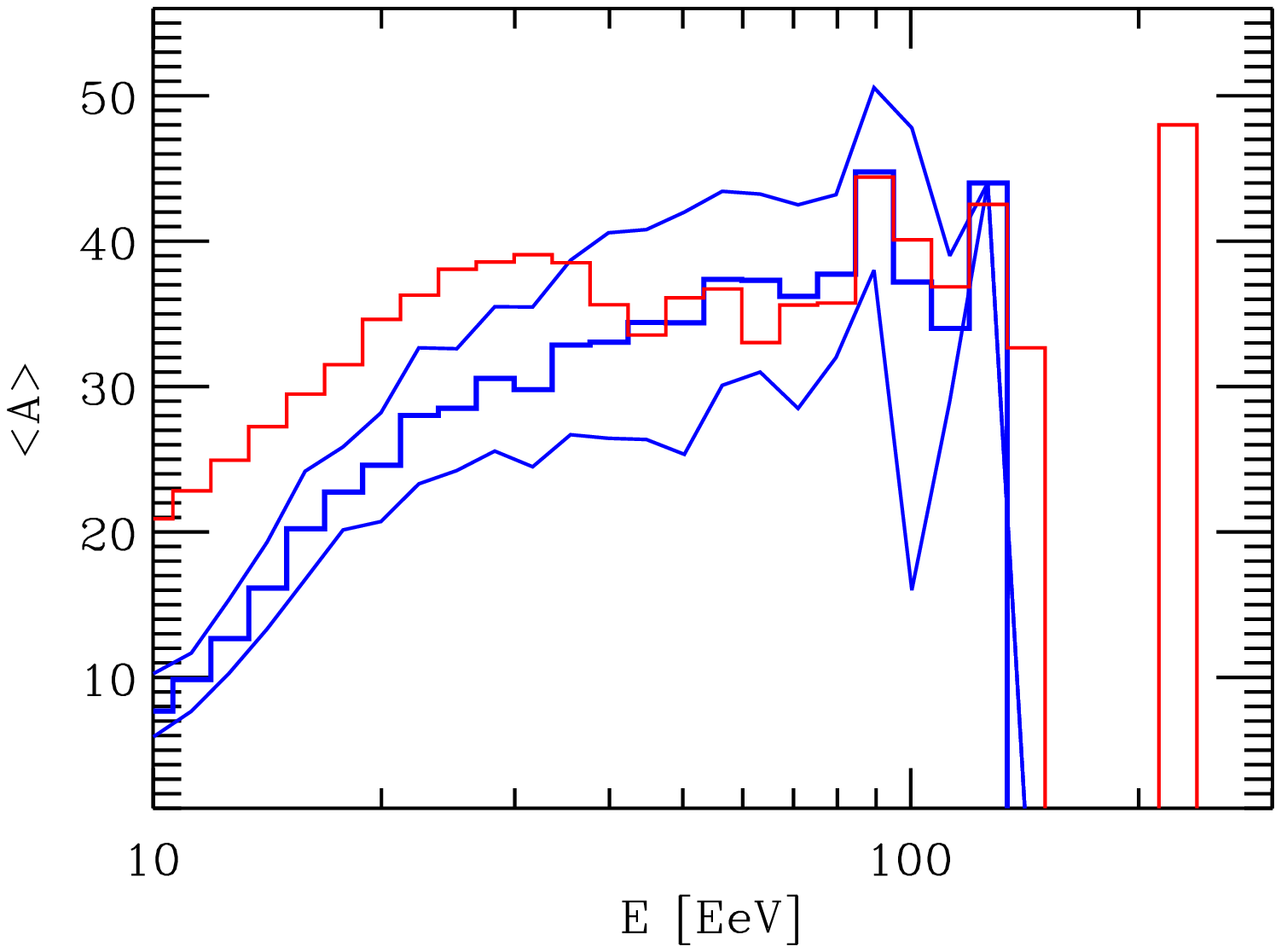}
\caption{Observed UHECR spectrum (top) and mean mass composition (bottom) versus energy
$E$ (1 EeV $\equiv 10^{18}$ eV)
from cluster accretion shocks for the case of $\alpha=1.7$ and $\beta=0.5$,
compared with the current data for HiRes (bars) and Auger (asterisks).
The histograms are the average result over different model realizations
for the cases with (thick) and without (thin) extragalactic magnetic fields,
and the thin curves outline the median deviations due to cosmic variance for the former case only.
The straight line in the top panel denotes the injection spectrum.}
\label{fig:speccomp}
\end{figure}

{\it Results and Discussion.}
Fig.\ref{fig:speccomp} displays our results for the observed spectrum and composition
for the case of $\alpha$=1.7 and $\beta$=0.5.
The spectrum is consistent with the current data for the HiRes \cite{Abb04}
and the Pierre Auger experiments \cite{Yam07}
within about 2 sigma of combined cosmic and statistical variance,
for both the differential and integral fluxes.
The latest analysis of the AGASA data \cite{Tak98} (not plotted)
may possibly result in similar spectra
depending on the choice of air shower simulations \cite{Tes06}.
Note that values of $\alpha <2$ are naturally expected at the high energy end for 
nonlinear shock acceleration that accounts for the dynamical back reaction from CRs \cite{MD01}.
For $\beta$, acceptable fits to the spectrum are obtained in the range 0.3-0.5
corresponding to a factor of $\sim$3-7 enhancement for Fe nuclei.
It remains to be seen whether this can be explained from first principles
in the context of nonlinear acceleration theories \cite{BK99}.
We predict a clear steepening in the spectra $\gtrsim 10^{20}$ eV,
due not only to photonuclear losses during propagation,
but also because our estimated $E_{\max}$ at the source is not far from this value.

Normalizing to the observed flux, the required source power above $10^{19}$ eV is inferred to be in the range
$P_{>19} \simeq (1-50) \times 10^{38} {\rm erg\ s^{-1} Mpc^{-3}}$ for the cases with EGMF,
fluctuating by a large factor depending on the specific realization of the source locations
with different propagation lengths to the observer.
Comparing with $P_{\rm acc}$ estimated above,
we deduce a reasonable range of $f_{\rm CR} \simeq 0.005-0.3$.
We note that for $\alpha<2$ as expected in nonlinear shock acceleration,
the energetics is dominated by CRs at the high energy end,
while those at lower energies do not contribute much power.
For the cases without EGMF,
$P_{>19} \simeq 3 \times 10^{37} {\rm erg\ s^{-1} Mpc^{-3}}$,
quite independent of the realization, leading to $f_{\rm CR} \simeq 0.002$.
Low values of $f_{\rm CR}$ could reflect inefficient escape from the system rather than inefficient acceleration.

Consistent with claims from HiRes \cite{Abb05},
the flux below $\simeq 2 \times 10^{19}$ eV is dominated by lighter nuclei.
This is due both to their prevalence at lower 
energies in the source spectrum as described above,
and to pileup of photodisintegrated fragments coming down
from heavier nuclei at higher energies during propagation.
The latter effect is more prominent with EGMF because of the longer propagation lengths.
The rapid increase in the average mass composition above $\simeq 2 \times 10^{19}$ eV,
a natural consequence of higher $Z$ nuclei extending to higher $E_{\max}$,
is a definitive prediction of our scenario to be verified by the new generation experiments.

\begin{figure}
\includegraphics[width=0.41\textwidth]{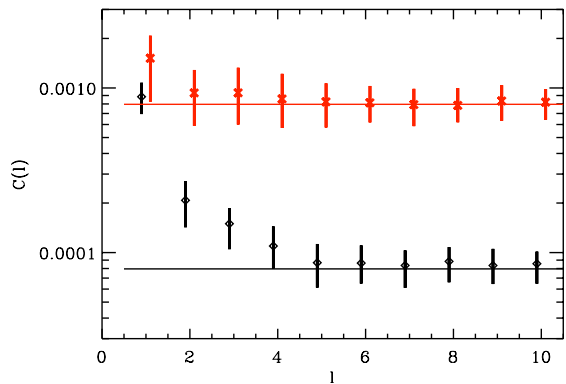}
\caption{Angular power spectrum $C(l)$ of UHECR arrival directions above $4 \times 10^{19}$ eV
versus multipole $l$,
for a realization with extragalactic magnetic fields and a single, dominant cluster at $D \sim$ 50 Mpc.
The crosses are for 100 events with AGASA + SUGAR exposure
and diamonds for 1000 events with Auger North + South exposure.
Vertical bars indicate statistical errors.}
\label{fig:aniso}
\end{figure}

Since massive clusters are relatively rare in the local universe within $\sim$300 Mpc,
the UHECR flux should be dominated by only a few nearby sources.
Nevertheless, strong deflections of the highly charged nuclei in EGMF
allow consistency with the observed global isotropy of UHECRs with the current statistics.
On the other hand, we predict that with a sufficient number of accumulated events,
significant anisotropy should appear toward a small number of individual sources.
Fig.\ref{fig:aniso} compares the angular power spectrum of arrival directions
expected for 100 and 1000 events above $4 \times 10^{19}$ eV,
in the case of a particular realization with a single, dominant cluster at distance $D \sim$ 50 Mpc,
somewhat resembling Perseus or Coma.
The sky distribution may show an excess of events extended by few tens of degrees,
offset from the true position of the cluster by a similar amount.
Such trends may in principle reveal important information on the EGMF.
Note, however, that these inferences could be affected when Galactic magnetic fields are included.
For the case with no EGMF, the anisotropy is too strong to be compatible with the current data,
although including weak EGMF and/or Galactic fields may alleviate this \cite{TYS06,CMS07}.
At any rate, a general feature of scenarios with heavy nuclei as UHECRs
is that a small number of powerful sources can be consistent
with the current global isotropy and lack of association with known objects.
This is in contrast to AGN scenarios with protons
that must rely on sources more numerous than powerful FRII radio galaxies,
since the latter are as rare as massive clusters
despite being the most promising AGN candidates \cite{RB93}.

A good match to the spectra can also be obtained without the $A^{\beta}$ factor if $\zeta \sim$0.5-1.0,
but such high metallicities may not be expected for the accreting gas \cite{DM01,CO06}.
The results are fairly insensitive to variations in $\alpha$ alone,
since prescribing the source injection composition at fixed $E/A$
implies a weighting factor $A^{\alpha-1}$ at a given $E$ \cite{SA05},
which counteracts the $E^{-\alpha}$ dependence.
On the other hand, a modest increase of $E_{\max}/Z$ by a factor of 2
produces a large spectral bump at a few times $10^{19}$ eV from photodisintegration products,
at odds with the observations.
Thus it is crucial that the source spectrum cuts off near the photodisintegration threshold on the FIRB,
which is naturally expected in our cluster scenario (Fig.\ref{fig:tacc}),
but is not obvious for nuclei acceleration in other contexts, e.g. AGNs \cite {RB93} or GRBs \cite{APO05}.

Although most previous studies of UHECR nuclei concentrated on the propagation aspects
\cite{SS99,ASM05,APO05},
here we have presented a physically plausible scenario of UHECR nuclei production by cluster accretion shocks
that can account for the current observations.
Combined with complementary information from X-ray and gamma-ray observations \cite{IAS05,ASM06},
detailed measurements of UHECR composition and anisotropy
with facilities such as the Pierre Auger Observatory,
the Telescope Array, and the future Extreme Universe Space Observatory
should provide a clear test of whether the largest bound structures in the universe
are also the largest and most powerful particle accelerators.

After this paper was submitted, the Auger collaboration announced new observational results
regarding the composition \cite{Ung07} and anisotropy \cite{Aug07} of UHECRs.
Our prediction of heavy-dominance at the highest energies
is concordant with the composition estimates,
whereas it could be disfavored by the anisotropy results.
Although the implications of these apparently conflicting results are not yet clear,
it is certainly possible that some fraction of UHECRs are heavy nuclei,
which can be accounted for by the scenario discussed in this paper.

{\it Acknowledgements.}
We thank V. Zirakashvili for inspiring discussions,
and P. Biermann, L. Drury, T. Jones, H. Kang,
J. Kirk, M. Nagashima, D. Ryu and M. Teshima for valuable communications.
F.M. acknowledges support by ETH-Z\"urich
through a Zwicky Prize Fellowship.


\end{document}